%Paper: hep-th/9403112
%From: Jose Mourao <mourao@phys.psu.edu>
%Date: Fri, 18 Mar 1994 16:09:26 -0500 (EST)

\baselineskip=12pt
\magnification=1000
%\input epsf
%\epsffile{/home/psuphys1/pullin/tex/header.ps}
\def\a{\alpha}
\def\Lapop{\displaystyle{{\hbox to 0pt{$\sqcup$\hss}}\sqcap}}
\def\b{\beta}

\def\e{\epsilon}

\def\l{\lambda}

\def\L{\Lambda}

\def\R{{\rm I\!R}}
\def\N{{\rm I\!N}}

\def\Q{{\mathchoice
{\setbox0=\hbox{$\displaystyle\rm Q$}\hbox{\raise 0.15\ht0\hbox to0pt
{\kern0.4\wd0\vrule height0.8\ht0\hss}\box0}}
{\setbox0=\hbox{$\textstyle\rm Q$}\hbox{\raise 0.15\ht0\hbox to0pt
{\kern0.4\wd0\vrule height0.8\ht0\hss}\box0}}
{\setbox0=\hbox{$\scriptstyle\rm Q$}\hbox{\raise 0.15\ht0\hbox to0pt
{\kern0.4\wd0\vrule height0.7\ht0\hss}\box0}}
{\setbox0=\hbox{$\scriptscriptstyle\rm Q$}\hbox{\raise 0.15\ht0\hbox to0pt
{\kern0.4\wd0\vrule height0.7\ht0\hss}\box0}}}}
\def\C{{\mathchoice
{\setbox0=\hbox{$\displaystyle\rm C$}\hbox{\hbox to0pt
{\kern0.4\wd0\vrule height0.9\ht0\hss}\box0}}
{\setbox0=\hbox{$\textstyle\rm C$}\hbox{\hbox to0pt
{\kern0.4\wd0\vrule height0.9\ht0\hss}\box0}}
{\setbox0=\hbox{$\scriptstyle\rm C$}\hbox{\hbox to0pt
{\kern0.4\wd0\vrule height0.9\ht0\hss}\box0}}
{\setbox0=\hbox{$\scriptscriptstyle\rm C$}\hbox{\hbox to0pt
{\kern0.4\wd0\vrule height0.9\ht0\hss}\box0}}}}

\font\fivesans=cmss10 at 4.61pt
\font\sevensans=cmss10 at 6.81pt
\font\tensans=cmss10
\newfam\sansfam
\textfont\sansfam=\tensans\scriptfont\sansfam=\sevensans\scriptscriptfont
\sansfam=\fivesans
\def\sans{\fam\sansfam\tensans}
\def\Z{{\mathchoice
{\hbox{$\sans\textstyle Z\kern-0.4em Z$}}
{\hbox{$\sans\textstyle Z\kern-0.4em Z$}}
{\hbox{$\sans\scriptstyle Z\kern-0.3em Z$}}
{\hbox{$\sans\scriptscriptstyle Z\kern-0.2em Z$}}}}
\def\semi{\bigcirc\kern-1em{s}\;}

% Spezielle Definitionen

%   Non-character macros

\newcount\foot
\foot=1
\def\note#1{\footnote{${}^{\number\foot}$}{\ftn #1}\advance\foot by 1}

\def\frac#1#2{{#1\over #2}}
\def\text#1{\quad{\hbox{#1}}\quad}

% Font-Definitions

\font\ftn=cmr8 scaled\magstephalf

\font\it=cmti10 scaled\magstephalf
\font\bf=cmbx10 scaled\magstephalf

\topskip 2cm
\def\agb{\overline {{\cal A}/ {\cal G}}}

\vbox{\rightline {CGPG-94/3-1}}

\centerline{\bf ON THE SUPPORT OF THE }
\centerline{\bf ASHTEKAR-LEWANDOWSKI MEASURE}

\vskip 2cm

\centerline{Donald Marolf and Jos\'e M. Mour\~{a}o$^\dagger$}

\centerline{\it Department of Physics}
\centerline{\it Center for Gravitational Physics and Geometry}
\centerline{\it The Pennsylvania
State University, University Park, PA 16802-6300}

\vskip 2cm

\centerline{\bf Abstract}

We show that the Ashtekar-Isham extension
$\overline {{\cal A} / {\cal G}}$ of the configuration space of
Yang-Mills theories $ {{\cal A} / {\cal G}}$
is (topologically and measure-theoretically)
the projective limit of a family of finite dimensional spaces
associated with arbitrary finite lattices.

These results are then used to prove that
${{\cal A} / {\cal G}}$ is contained in a zero measure
subset of $\overline {{\cal A} / {\cal G}}$ with respect to the
diffeomorphism invariant  Ashtekar-Lewandowski measure
on $\overline {{\cal A} / {\cal G}}$.  Much as in scalar field theory,
this implies that states in the quantum theory associated with this
measure can be realized as functions on the ``extended"
configuration space $ {\overline {{\cal A} / {\cal G}}}$.

\vfill

\item{}
$^\dagger${\it On leave of absence from Dept. F\'{\i}sica, Inst. Sup.
T\'{e}cnico, 1096 Lisboa,
PORTUGAL}

\eject

 {\bf 1 Introduction}

\vskip 0.5cm

The usual canonical approach to quantization of a (finite dimensional)
system
defines states as functions on a configuration space and defines an
inner product  of two such functions $\psi$ and $\phi$ through
$$
(\psi,\phi) = \int_{\cal Q} d\mu \ \psi^* \phi
$$
where $\mu$ is some measure on the configuration space $\cal Q$.
Naively applying this procedure to Yang-Mills theories produces a
``connection representation" with states that are functions of the
Yang-Mills connection.  In particular, these states are functions on
the quotient space ${\cal A}/{\cal G}$, where ${\cal A}$ is the
space of ($C^1$-)connections and ${\cal G}$ is the group of ($C^2$-)gauge
transformations. The same is true for gravity
formulated in terms of Ashtekar variables before one imposes the diffeomorphism
and hamiltonian constraints [1,2].

A more sophisticated
analysis of examples, such as scalar field theory [3-5], shows that the
domain
space of the wave functions may not be exactly the classical configuration
space. Instead, some extension of ${\cal Q}$ is required.

In
order to define an inner
product for a connection representation, one expects to give
${\cal A} / {\cal G}$, or some suitable extension, the
structure of a measurable space (by choosing the measurable sets)
and to define appropriate measures.
Ashtekar and Isham described an algebraic program to
construct such measures in [2].
They proposed, for a compact gauge group $G$,  a
compact extension  $\overline
{{\cal A} / {\cal G}}$ of ${\cal A} / {\cal G}$ on which regular Borel
measures are well defined and are in one-to-one correspondence
with positive continuous linear  functionals on a
certain $C^*$-algebra of connection observables known as  the holonomy algebra
$\overline {\cal HA}$. In [6], Ashtekar and Lewandowski
constructed such a Borel measure
$\mu_{AL}$ on $\overline {{\cal A} / {\cal G}}$
that is both diffeomorphism invariant and
strictly positive on continuous cylindrical functions.
To do so they, and independently Baez in [7],
introduced the concepts of ``cylindrical sets" and ``cylindrical functions"
on  $\overline {{\cal A} / {\cal G}}$. Baez then generalized the
Ashtekar-Lewandowski measure by finding an infinite
dimensional space of diffeomorphism invariant measures.
In [6] it was also shown that the Ashtekar-Isham space
$\overline {{\cal A} / {\cal G}}$ is in one-to-one correspondence
with the set of homomorphisms
from the group of piecewise
analytic hoops
(i.e. based loops modulo an equivalence relation defined by the
holonomies)
${\cal HG}_{x_0}$
to the gauge group $K$, modulo conjugation.

In what follows, we reinterpret some of the results of
[2, 6] in terms of the theory of projective limits.  In
particular, we consider projective limits of infinite
families of finite dimensional topological and measurable spaces
associated with {\it arbitrary} finite lattices.
This theory provides an appropriate framework for studying
different properties of $\overline {{\cal A} / {\cal G}}$, both
from
the topological and measure theoretical points of view.
Our main result is the use of this formalism to prove that
the space $ {{\cal A} / {\cal G}}$ is contained  in a zero
measure subset of $\overline {{\cal A} / {\cal G}}$ (with respect
to the Ashtekar-Lewandowski measure).

The present work is organized as follows. In Sect. 2 we recall
(mainly from [8]) some aspects of the theory of projective limits
of infinite families of measurable spaces. Sect. 3 is devoted to
reinterpreting some results of [2, 6] in the language of
projective limits. In particular
we show that $\overline {{\cal A} / {\cal G}}$
is a projective limit of a family of finite-dimensional
spaces and that  the Gel'fand topology on the spectrum
$\overline {{\cal A} / {\cal G}}$ coincides with the
Tychonov topology on the projective limit.  While the
Ashtekar-Isham space ${\overline {{\cal A}/{\cal G}}}$ is defined
only for compact gauge groups $G$, the projective limit is defined for
the noncompact case as well.
On the
measure theoretical side we show that the measurable space
 $(\overline {{\cal A} / {\cal G}}, {\cal B}(\overline {\cal C}))$
(where ${\cal B}(\overline {\cal C})$ denotes
 the minimal $\sigma$-algebra
containing the cylindrical sets $\overline {\cal C}$) is isomorphic to the
projective
limit. In Sect. 4 we prove the main result of the paper stated above.
Sect. 5 is devoted to the study of the additive, but not
$\sigma$-additive, measure $\hat \mu_{AL}$ induced
by $\mu_{AL}$ on the (finite) algebra $\cal C$ of cylindrical sets of
$ {{\cal A} / {\cal G}}$
$$
{\cal C} = \{ \bar C \cap  {{\cal A} / {\cal G}} \ ; \quad \bar C \subset
\overline {\cal C} \}
$$
where $\overline {\cal C}$ denotes the algebra of cylindrical sets
on $\overline {{\cal A} / {\cal G}}$. We show that $\hat \mu_{AL}$
cannot be extended to a $\sigma$-additive measure on
${{\cal A} / {\cal G}}$ and that the space
of square integrable (cylindrical) functions on
 ${{\cal A} / {\cal G}}$ is not complete.
We also prove that the Cauchy completion
of this space is
$L^2 \left( \overline {{\cal A} / {\cal G}},
 \mu_{AL}, {\cal B}(\overline {\cal C})\right)$,
justifying the use of the
``generalized connections" in ${\overline {{\cal A}/{\cal G}}}$.

\vskip 1.5cm

\vbox{

{\bf 2. Projective limit measurable spaces}}

\vskip 0.5cm

In the present section we recall, mainly from [8], the relevant aspects
of a class of measures on infinite dimensional spaces
which are obtained as rigorously defined limits of measures
on finite dimensional spaces. This class contains the direct product measures
(on $\R^\infty$ for example) and the projective limit measures.  First,
however, we introduce some more terminology and
notation that will prove useful.

The pair $(X, {\cal B})$ (or $(X, {\cal F})$), where $X$ is a set
and $\cal B$ (${\cal F}$) is a $\sigma$-algebra (algebra)
of subsets of $X$, will be called a $\sigma$-measurable (measurable) space.
In the mathematical literature, definitions of a measurable space have been
given both that require $\cal B$ to be a $\sigma$-algebra  and that
require only that ${\cal B}$ be closed under finite operations.
As  we will be interested in a comparison of these
two cases it will be convenient to use the above terminology to
distinguish between them.

We will be interested in $\sigma-additive $ probability measures
on $\cal B$, which are, by definition non-negative, normalized and
$\sigma$-additive functions on the $\sigma$-algebra $\cal B$.
That is, such a measure $\mu$ satisfies:
$$
\eqalignno{
\mu(B) &\geq 0 \quad , \quad B \in {\cal B}    & (2.1a) \cr
\mu(X) & =  1 \ ,     & (2.1b) \cr
\mu(\cup_{i=1}^{\infty} B_i) &= \sum_{i=1}^\infty \mu( B_i) \qquad ,
\quad B_i \in {\cal B}, \quad B_i \cap B_j = \emptyset , \ \ i \neq j
\ .   & (2.1c) \cr}
$$
Additive measures on an algebra $\cal F$
satisfy (2.1) with $\cal B$ replaced by $\cal F$ and with only
finite unions and sums in (2.1c). For a given measure $\mu$
on $\cal F$, an important question is whether or not
it can be extended to a $\sigma$-additive measure on
${\cal B}({\cal F})$, the minimal $\sigma$-algebra
that contains $\cal F$. A necessary and sufficient condition
for extendibility is given by the Hopf theorem [8]:

\bigskip

\vbox{

\item{} {\bf Theorem 2.1} (Hopf Theorem)

\item{} A measure $\mu$ on $\cal F$ can be extended to a $\sigma$-additive
measure on ${\cal B}({\cal F})$ if and only if for every decreasing sequence
$\{F_i\}$ such that
$F_i  \in {\cal F} \ , \ F_1 \supset \ldots \supset F_n \supset
\ldots$ \ with $\cap_{i=1}^{\infty} F_i = \emptyset$, we have
$$
\lim_{i \rightarrow \infty} \mu (F_i) = 0   \ .   \eqno(2.2)
$$
}

\vskip .5cm

{\noindent}Essentially, the condition (2.2) allows an extension $\tilde \mu$
to be consistently defined on elements of ${\cal B}({\cal F})$ as
limits of $\mu$-measures of sets in $\cal F$. The triplet
$\{(X, {\cal B}), \mu \}$  ($\{(X, {\cal F}), \mu \})$,
where $\cal B$ ($\cal F$) is a $\sigma$-algebra (algebra)
and $\mu$ is  $\sigma$-additive (additive) is called
a $\sigma$-measure (measure) space.

The possibility of extending a measure $\mu$ on $\cal F$
to a $\sigma$-additive measure $\tilde{\mu}$ on ${\cal B}({\cal F})$
is in particular relevant to physical applications
in quantum mechanics. Recall that quantum mechanical systems are
often defined by first giving a linear pre-Hilbert space
and then completing this space with respect
to an inner product. In general, if $\mu$ is cylindrical  but not
$\sigma$-additive, the space ${\cal H}$
of $\mu$-square integrable cylindrical functions on $X$
(denoted through $ {\cal C}L^2 (X, {\cal F}, \mu)$)
 is only a pre-Hilbert space.  Such spaces will be discussed
in section 5.  However,
if $\mu$ is extendible to a
$\sigma$-additive measure $\tilde \mu$ on $(X, {\cal B}({\cal F}))$
then the Cauchy
completion of $\cal H$ leads  to the space $\tilde {\cal H} =
 L^2 \left(X, {\cal B} ({\cal F}), \tilde \mu \right)$ (see section 5).
On the other hand if $\mu$ is not extendible
then the  Cauchy completion of $ {\cal C}L^2 (X, {\cal F}, \mu)$ leads in
general to a space with state-vectors which cannot be expressed
as functions on the initial space $X$.
This is the case in scalar field theory if one considers
$X = {\cal S}(\R^3)$ (the Schwarz space of rapidly
decreasing smooth $C^\infty$ functions on $\R^3$) and $\mu$ is a cylindrical
measure defined with the help of a positive definite
function on ${\cal S}(\R^3)$, continuous in the nuclear
space topology (see [3, 5, 8]). As we shall see in Sect. 5 this is
also the case in Yang-Mills theory if we take
${\cal H} =  {\cal C}L^2 ({\cal A} / {\cal G}, {\cal F}
= {\cal C}, \hat \mu_{AL})$,
where $\hat \mu_{AL}$ is the Ashtekar-Lewandowski
measure on ${\cal A} / {\cal G}$. In the scalar field case
the Cauchy completion of $  {\cal C}L^2 ({\cal S}(\R^3), {\cal F}, \mu)$
gives the space of square integrable functions on ${\cal S}'(\R^3)$
(the space of tempered distributions), while in the Yang-Mills case
the completion of $  {\cal C}L^2 ({\cal A}/ {\cal G}, {\cal C}, \hat \mu_{AL})$
gives the space
$ L^2 (\overline {{\cal A}/ {\cal G}},
{\cal B}({\overline {\cal C}}), \mu_{AL})$
of square integrable functions on the Ashtekar-Isham space
$\overline {{\cal A} / {\cal G}}$ of generalized ``distributional"
connections modulo gauge transformations.

Let $\{(X, {\cal B}), \mu \} $ be a $\sigma$-measure space.
The subset $Y \subset X$ is said to be  $\mu$-thick in
$X$ if for every $B \in {\cal B}$ such that $B \cap Y = \emptyset,
 \quad \mu(B) = 0$. If $Y$ is $\mu$-thick in $X$ then $\mu$
induces a $\sigma$-additive measure $\mu_Y$ on the $\sigma$-measurable space
$$
(Y, {\cal B}_Y)   \ , \eqno(2.3a)
$$
where ${\cal B}_Y = \{B \cap Y \ , \ B \in {\cal B} \}$,
through
$$
\mu_Y (B \cap Y) = \mu (B) \ , \quad \forall B
\in {\cal B} \ .\eqno(2.3b)
$$
The measure $\mu_Y$ is called the trace of the measure $\mu$ on
$Y$ [8]. If $Y$ is not $\mu$-thick on $X$ then (2.3b) is
not well defined.

Note that if $Y$ is $\mu$-thick in $X$ then
$$
L^p(X, \mu, {\cal B}) \cong L^p(Y, \mu_Y, {\cal B}_Y)
$$
so that if we are concerned only with such  spaces we can restrict
ourselves to $Y$ and $\mu_Y$. This is particularly convenient
when the set $Y$ has advantages (for instance from the
``differentiable" point of view) over $X$.
When a set $Y$ is $\mu$-thick in $X$ we say that the support
of the measure $\mu$ is contained in $Y$. An illustrative
example is the one given by the Wiener measure on $\R^{[0,1]}$
used in the (euclidean) path integral formulation of quantum mechanics.
In this case the support of the measure is contained in the
space  $Y = C^0 ([0,1])$ of continuous functions on the interval [3].

The inclusion map from $Y$ to $X$ above is referred to as measurable.  In
general, a map between $\sigma$-measurable (measurable) spaces
$$
\phi : \quad X_1 \rightarrow X_2   \eqno(2.4)
$$
is called measurable if for every measurable set $B_2 \in {\cal B}_2$
the set $\phi^{-1}(B_2)$ is measurable i.e. $\phi^{-1}(B_2) \in
{\cal B}_1$. $\phi$ in (2.4) is called an isomorphism of
$\sigma$-measurable (measurable) spaces if it is bijective and if
both $\phi$ and $\phi^{-1}$ are measurable.

Let us now briefly review (see [8]) the construction
of infinite products of $\sigma$-measurable spaces and of (projective) limits
of infinite projective families of $\sigma$-measurable spaces. Let
$$
\{(X^{(\lambda)}, {\cal B}^{(\lambda)} )\}_{\lambda \in \Lambda}
\eqno(2.5)
$$
be an indexed
family of $\sigma$-measurable spaces. The product $\sigma$-measurable
space $(X^{(\Lambda)}, {\cal B}^{(\Lambda)})   $
is, by definition, given by
$$
X^{(\Lambda)} = \prod_{\lambda \in \Lambda} X^{(\lambda)}     \eqno(2.6)
$$
with ${\cal B}^{({\L})}$ being the minimal $\sigma$-algebra for which all
the projections
$$
\eqalign{
p_{\lambda_0} \ : X^{(\Lambda)} &\rightarrow  X^{(\lambda_0)} \cr
(x_\lambda)_{\lambda \in \Lambda} & \mapsto x_{\lambda_0} \cr} \eqno(2.7)
$$
are measurable. That is, ${\cal B}^{({\L})}$
 is the $\sigma$-algebra  generated by
the inverse images of measurable  sets in $X^{(\lambda)}$ under
 the projections $p_\lambda$. If all the $X^{(\lambda)}$,
$\lambda \in \Lambda$  are different copies of the same set
$Y$ with the same $\sigma$-algebras ${\cal B}^{(\lambda)} = {\cal B}$
then the points of $X^{(\Lambda)} = Y^{\Lambda}, \ x \in Y^\Lambda$ are
(arbitrary) maps from $\Lambda$ to $Y$:
$$
\eqalign{
 x \in X^{(\Lambda)} &=  Y^\Lambda  \Leftrightarrow x :  \Lambda \rightarrow Y
\cr
x &= (x_\lambda)_{\lambda \in \Lambda}  \ ; \quad
x_{\l} \in Y  \cr}  \eqno(2.8)
$$
Examples are the set of all sequences of real numbers
$$
\R^\infty = \prod_{j \in \N } \R_{(j)}          \qquad , \
(\R_{(j)} = \R) \eqno(2.9)
$$
and the set of all real valued functions on the interval $[0, 1]$
$$
\R^{[0, 1]} = \prod_{t \in [0, 1]} \R_t \qquad , \  (\R_t = \R)  \eqno(2.10)
$$

Suppose we have a $\sigma$-additive measure $\mu$ on ${ X}^{(\Lambda)}$
in (2.6). Let $\cal L$ be the family of all finite subsets of $\Lambda$
and for $L \in {\cal L}$ let
$(X^{(L)}, {\cal B}^{(L)})$
 be the partial products of $\sigma$-measurable spaces
with
$$
X^{(L)} = \prod_{\lambda \in L} X^{(\lambda)}   \eqno(2.11)
$$
and the corresponding ${\cal B}^{(L)}$. Then all the projections
$$
\eqalign{
p_L : \quad X^{(\Lambda)} &\rightarrow X^{(L)}   \cr
p_L((x_\lambda)_{\lambda \in \Lambda}) &= (x_\lambda)_{\lambda \in L} \cr}
\eqno(2.12)
$$
are measurable. Consider the family $\{ \mu_L \}_{L \in \cal L}$ of
$\sigma$-additive measures on $X^{(L)}$ defined by
the pushforwards of the measure $\mu$
$$
\mu_L(B) = \mu(p_L^{-1}(B)) \eqno(2.13)
$$
for $B \in {\cal B}_L$, which, in the notation of measure theory
is written as
$$
\mu_L = (p_{L})_*  \mu.
$$
This family satisfies a self consistency condition:
$$
L \subset L' \ \Rightarrow \ \mu_L = (p_{LL'})_*   \mu_{L'} \ ,  \eqno(2.14)
$$
where $p_{LL'}$ denote the measurable
projections from $X^{(L')}$ to $X^{(L)}$.
In [8] (see corollary to Th. 10.1) are found conditions for which the
converse is also true:

\vskip 0.5cm

\vbox{

\item{} {\bf Proposition 2.2}

\item{} For a family of $\sigma$-compact or complete and separable,
metric spaces every family of Borel measures
that is consistent in the sense of (2.14)
can be extended to a $\sigma$-additive measure
on the product $\sigma$-measurable space.}

\vskip 1cm

Such a measure is in fact defined by (2.13), i.e. for $B_L \in {\cal B}_L$,
$\mu(p_L^{-1}(B_L))$ is defined to be just $\mu_L(B_L)$.
Recall that a topological space $(X, \tau)$ is said
to be $\sigma$-compact if $X$ can be represented as a countable union
of compact sets.

Notice ([8]) that a measure $\mu$
satisfying (2.13) and given on the algebra
$$
{\cal F}^{({\cal L})} = \cup_{L \in \cal L} \ p_L^{-1}({\cal B}_L)
\eqno(2.15) $$
always exists. The only question is whether $\mu$ can be extended
to a $\sigma$-additive measure $\tilde \mu$ on ${\cal B}^{({\cal L})}$,
which
is the minimal $\sigma$-algebra that contains (2.15)
$$
{\cal B}^{({\cal L})} =
{\cal B} ( \cup_{L \in \cal L} \ p_L^{-1} ({\cal B}_L))
\ . \eqno(2.16)
$$
For instance, in the example of $X^{({\L})} = \R^{[0,1]}$ the question
is to know when a self-consistent family of measures
$\{ \mu_{t_1, \ldots , t_n} \}_{t_1, \ldots , t_n \in [0,1]}$
on the finite dimensional spaces
$$
\R^n = \prod_{i=1}^n \R_{t_i}    \eqno(2.17)
$$
defines a $\sigma$-additive measure on the infinite
dimensional space
$$
\R^{[0,1]}   \ . \eqno(2.18)
$$
Quite remarkably, in this example and in many others relevant
to quantum field theory the answer is  affirmative, as indicated by
Proposition 2.2.

An infinite product $\sigma$-measure space can also be realized
as a ``projective limit" (which we will define next).  However, the
product space $X^{(\Lambda)}$ is a projective
limit not of the family of spaces $X^{(\lambda)}$ labelled by
$\lambda \in  \Lambda$ but rather of the family
of spaces $X_L = X^{(L)}$ labelled by $L \in {\cal L}$, the set of
all finite subsets of $\Lambda$.  In general, a projective limit
space can be defined for any
``projective family" of $\sigma$-measurable spaces; that is, for any
family
$$
\{(X_L, {\cal B}_L), p_{LL'} \}_{L,L' \in \cal L} \ , \eqno(2.19)
$$
of the following form.  The set $\cal L$ is taken to be
directed, i.e. partially ordered and
such that for any two elements
$L_1, L_2 \in {\cal L}$, there is some $L $ such
that $ L_1 \leq L$
and $L_2 \leq L$. We will also assume that $\cal L$ does not
have a maximum. Here $p_{LL'}$ are measurable projections,
i.e. surjective mappings
$$
p_{LL'} \ : \ X_{L'} \rightarrow X_L \quad L < L'  \eqno(2.20a)
$$
satisfying
$$
p_{LL'} \circ p_{L'L''} = p_{LL''} \quad {\rm for}  \ L < L' < L''
\ . \eqno(2.20b)
$$

Now, let $(X^{({\cal L})}, {\cal B}^{({\cal L})})$
denote the direct product of the family
$\{(X_L, {\cal B}_L)\}_{L \in \cal L}$
$$
X^{({\cal L})} = \prod_{L \in \cal L} X_L  \ .
$$
Then the projective limit of the family (2.19)
is by definition the $\sigma$-measurable space
$(X_{\cal L}, {\cal B}_{\cal L})$ , where
$$ \eqalignno{
 X_{\cal L}  \subset X^{({\cal L})}  \quad ; \qquad
X_{\cal L}  & = \lbrace (x_L)_{L \in {\cal L}} \in X^{({\cal L})}
: \ L < L' \Rightarrow
x_L = p_{LL'} (x_{L'}) \rbrace  & (2.21a)   \cr
\noalign{\hbox{and}}
{\cal B}_{\cal L} &= \lbrace B \cap X_{\cal L}   \ : \ B \in {\cal
B}^{({\cal L})}
 \rbrace \ .   & (2.21b)    \cr}
$$
That is, $X_{\cal L}$ is the subset of $X^{({\cal L})}$ that
is consistent with the projections $p_{LL'}$.
Note that a direct product space can also be thought of as
a projective limit of
the spaces formed by taking arbitrary finite products of the
factors. A family of measures $(\mu_L)_{L \in \cal L}$ is said
to be self-consistent if it satisfies (2.14) with $L \subset L'$
replaced by $L < L'$. A measure $\mu$ on ${\cal B}_{\cal L}$ always
defines a self consistent family of measures $(\mu_L)_{L \in \cal L}$
through (2.13) and a consistent family $(\mu_L)_{L\in {\cal L}}$ defines
a finitely additive measure on $X_{\cal L}$ through (2.13) as well.
A measure on $X_{\cal L}$ defined by such a family is called
cylindrical.   An important result  is
(see  [8] Corollary to Th. 10.1):

\vskip 0.5cm

\vbox{

\item{} {\bf Proposition 2.3}

\item{} Under the same conditions as in Proposition 2.2, a self-consistent
family of Borel measures on a projective family (2.19)
defines a cylindrical measure that
can be extended to a $\sigma$-additive measure in the
projective limit
$\sigma$-measurable space (2.21) if for every increasing sequence}
$$
{\cal M}  = \{L_i\}_{i=1}^\infty \subset {\cal L} \ : \
L_1 < L_2 < \ldots < L_n < \ldots
$$
\item{} with projective limit
 $(X_{\cal M}, {\cal B}_{\cal M})$
  the projection
$$
\eqalign{
p_{\cal M} \ : \ X_{\cal L} &\rightarrow X_{\cal M}   \cr
p_{\cal M}((x_L)_{L \in {\cal L}}) & =  (x_{L_n})_{L_n \in \cal M}   \cr}
$$
is surjective.

\vskip 1.5cm

\vbox{

{\bf 3. Ashtekar-Isham space $\overline {{\cal A} / {\cal G}}$
as a projective limit}
}

\vskip 0.5cm

Let $ {{\cal A} / {\cal G}}$ denote the space ${\cal A}$ of smooth
$C^1$
$G$-connections modulo the group ${\cal G}$ of gauge transformations
on a three dimensional analytic manifold $\Sigma$ where, as in [6],
the gauge group $G$ is
assumed to be $U(N)$ or $SU(N)$. Following
[6], we consider the $G$-hoop group ${\cal HG}_{x_0} =
{\cal L}\Sigma_{x_0} / \sim$ where
${\cal L}\Sigma_{x_0}$ is the space of piecewise analytic
loops based at $x_0$ (see [6]) and
the equivalence relation $\sim$ is
$$
\alpha , \beta \in {\cal L}\Sigma_{x_0} \ , \
\alpha \sim \beta \ \ \hbox{\rm if and only if} \ \  H(\alpha, A) =
H(\beta, A) \ , \ \forall A \in {\cal A} \ . \eqno(3.1)
$$
Here, $H(\alpha, A)$ denotes the holonomy corresponding to
the connection $A$ and the loop $\alpha$.
The Ashtekar-Isham space $\overline {{\cal A} / {\cal G}}$
is a ``compactification" of $ {{\cal A} / {\cal G}}$ obtained as follows
(see [2]). Let $T_\alpha \ , \alpha \in {\cal L}\Sigma_{x_0}$ denote
the Wilson loop function on ${{\cal A} / {\cal G}}$
defined by
$$
T_\alpha (A) = T_{[\alpha]}([A]) \equiv {1 \over N} Tr H(\alpha, A) \ .
\eqno(3.2) $$
where $[\alpha]$ denotes the equivalence class of $\alpha$ in
${\cal HG}_{x_0}$, $[A]$ denotes the equivalence class of $A$
in ${\cal A}/{\cal G}$
and the trace is taken in the fundamental representation
of the gauge group.
  In the following, for simplicity, $\alpha, \beta$ will denote
hoops.
The holonomy algebra $\widetilde {\cal HA}$ is the commutative
$C^*$-algebra generated by the Wilson loop functions.
The Ashtekar-Isham space $\overline {{\cal A} / {\cal G}}$ is the
compact Hausdorff space that is the
spectrum [2] of $\widetilde {\cal HA}$ in which
$ {{\cal A} / {\cal G}}$ is densely embedded [2,6,8].

Ashtekar and Lewandowski [6] obtained a useful
algebraic characterization of the space $\overline {{\cal A} / {\cal G}}$.
They proved that there is a one-to-one correspondence between
$\overline {{\cal A} / {\cal G}}$ and the space of {\it all}
 homomorphisms
from the hoop group ${\cal HG}_{x_0}$ to the gauge group $G$,
modulo conjugation. We will therefore identify
these two sets and write
$$
\eqalign{
\tilde h &= [h_0] \in \overline {{\cal A} / {\cal G}} \Leftrightarrow \cr
[h_0] &= \lbrace h
\in Hom({\cal HG}_{x_0} ,G) \ :
(h(\alpha))_{\alpha \in {\cal HG}_{x_0}} = (g h_0(\alpha) g^{-1})_{\alpha
\in {\cal HG}_{x_0}} \ , \ \hbox{\rm for some } g \in G \rbrace  \cr}
\eqno(3.3)
$$
where $g$ above does not depend on the hoop ${\alpha}$.
Notice that no continuity condition
has been imposed on the homomorphisms $h$ in (3.3). This will allow us
to interpret $\overline {{\cal A} / {\cal G}}$ (both topologically and
measure theoretically) as a projective limit of finite dimensional spaces.

Let $\cal L$ denote the set of all subgroups of
${\cal HG}_{x_0} $ generated by a finite number of hoops
$\beta_1 , \ldots , \beta_n$ that are strongly independent in the
sense of [6], i.e. such that
  loop representatives of the hoop equivalence
classes
 $\beta_i$ can be chosen in such a way that each
contains an open segment
which is traced exactly once and which intersects any of the
other representative loops
at most at a finite number of points.
 Then
$$
\eqalign{
S^* &\in {\cal L} \ \Leftrightarrow \cr
S^* & = \lbrace \hbox{\rm group generated by} \ \beta_1 , \ldots , \beta_n
\rbrace \subset {\cal HG}_{x_0}   \cr} \eqno(3.4)
$$
and we write $S^* = S^*[\beta_1 , \ldots , \beta_n ]$. Now let $H_{S^*}
= Hom(S^*,G)/Ad$
be the set of equivalence classes of homomorphisms from $S^*$ to $G$
under conjugation. If $S^* = S^*[\beta_1 , \ldots , \beta_n]$
then, as shown in [6], a homomorphism from $S^*$ to $G$ is known
if and only if we know it on the hoops $\beta_1, \ldots , \beta_n$
so that we
have the one-to-one correspondence
$$
\eqalignno{
H_{S^*} & \rightarrow G^n / Ad   & (3.5a) \cr
[h] & \mapsto [h(\beta_1), \ldots , h(\beta_n)] \ .  & (3.5b) \cr}
$$
Consider now the following projective family of finite dimensional spaces
$$
\lbrace (H_{S^*}), p_{S^*S^{*'}} \rbrace_{S^* , S^{*'} \in {\cal L}}
\ , \eqno(3.6)
$$
where $p_{S^*S^{*'}}$, $S^* \subset S^{*'}$, denotes the mapping
$$
\eqalignno{
p_{S^* S^{*'}} \ &: \ H_{S^{*'}} \rightarrow H_{S^*} & (3.7a)  \cr
p_{S^* S^{*'}}([h_{S^{*'}}]) &= [h_{S^{*'}} \mid_{S^*} ]  \   &(3.7b) \cr}
$$
and $h_{S^{*'}}\mid_{S^*}$ denotes the restriction of $h_{S^{*'}}$ to the
subgroup $S^*$ of $S^{*'}$.
{} From [6] we see that these projections are surjective.
According to (2.21) the projective limit $H_{\cal L}$ of the
family (3.6) is given by
$$
\eqalign{
H_{\cal L} &\subset H^{({\cal L})} = \prod_{S^* \in \cal L} H_{S^*}   \cr
H_{\cal L} &= \lbrace ([h_{S^*}])_{S^* \in \cal L} \in H^{({\cal L})}
 : \ S^* \subset
S^{*'} \Rightarrow [h_{S^*}] = p_{S^*S^{*'}} ([h_{S^{*'}}]) \rbrace \  . \cr}
\eqno(3.8)
$$
We will now show that this is just the Ashtekar-Isham space
$\overline {{\cal A}/{\cal G}}$.

\vskip 0.5cm

\vbox{

\item{}{\bf Proposition 3.1}
\item{} There is a bijective map $\phi$
$$
\eqalignno{
\overline {{\cal A} / {\cal G}} \ &{\buildrel \phi \over \rightarrow} \
H_{\cal L} &(3.9a) \cr
\noalign{\hbox{\qquad defined by}}
[h] & \mapsto ([h_{S^*}])_{S^* \in \cal L}   \ ; \ h_{S^*} = h\mid_{S^*}
\ .  & (3.9b) \cr}
$$
}

\vskip 0.5cm

\vbox{

\item{}{\bf Proof}

\item{} Consider the space $Hom({\cal HG}_{x_0}, G)$ of all homomorphisms
from ${\cal HG}_{x_0}$ to $G$ and the projective family
$\lbrace Hom(S^*,G), \tilde p_{S^*S^{*'}} \rbrace_{S^*,S^{*'} \in \cal L}$,
where $\tilde p_{S^*S^{*'}} \ : \  \tilde p_{S^*S^{*'}}(h_{S^{*'}}) =
h_{S^{*'}} \mid_{S^*}$, $ \
S^* \subset S^{*'}$ are surjective maps from $ Hom(S^{*'},G)$
to $ Hom(S^*,G)$. Let $K^{({\cal L})}$ be the infinite product
space and $K_{\cal L}$ be the projective limit space
of this family}
$$
K_{\cal L} = \lbrace (h_{S^*})_{S^* \in \cal L} \in K^{({\cal L})} \ :
\ S^{*} \subset S^{*'} \ \Rightarrow \
  h_{S^{*}} =  \tilde p_{S^*S^{*'}}(h_{S^{*'}}) \rbrace  \ . \eqno(3.10)
$$
\item{} We will need the following lemmas.

\vskip 0.5cm

\vbox{

\item{} {\bf Lemma 3.2}

\item{} The map
$$
\eqalign{
& \tilde \phi \  : \ Hom({\cal HG}_{x_0}, G) \ \rightarrow \ K_{\cal L} \cr
& \tilde \phi(h)  = (h \mid_{S^*})_{S^* \in \cal L}  \cr} \eqno(3.11)
$$
is bijective and $Ad$-equivariant, i.e. $Ad_g \circ \tilde \phi =
\tilde \phi \circ Ad_g$ for every $g \in G$. }

\vskip 0.5cm

\vbox{

\item{} {\bf Proof of Lemma 3.2}

\item{} The injectivity of $\tilde \phi$ is trivial. Let us prove
that $\tilde \phi$ is surjective. Fix an arbitrary element
$(h^0_{S^*})_{S^* \in \cal L} \in K_{\cal L}$. Let us construct
the homomorphism $h^0$ which is the pre-image of this element.
Let $\a$ be an arbitrary hoop and $S_1^* \in \cal L$ such that
$\a \in S_1^*$ ($S_1^*$ always exists for a piecewise
analytic hoop $\a$ [2]).
 Then choose $h^0(\a) =
h^0_{S^*_1}(\a)$.
To see that $h^0(\a)$ does not depend on the choice of the
finitely generated group $S^*_1 \ni \a$ let $\a \in \tilde S^*_1$
and $S^*_2$ be a subgroup which contains both
$S^*_1$ and $\tilde S^*_1$. Then, according to the definition
of $K_{\cal L}$, we have
$ h^0_{S^*_1}(\a) =  h^0_{S^*_2}(\a)$ and
$ h^0_{\tilde S^*_1}(\a) =  h^0_{S^*_2}(\a)$ which implies that
$ h^0_{\tilde S^*_1}(\a) =  h^0_{S^*_1}(\a)$.
We can easily show that $h^0$ constructed in this way is an homomorphism
and that the map $\tilde \phi$ is equivariant
$$
\hskip 7cm Q.E.D.
$$ }

\vskip 0.5cm

\item{} Lemma 3.2 implies that the map $\tilde \phi$
induces a bijective map $\phi_1$
$$
\eqalign{
 & \phi_1 \  : \ Hom({\cal HG}_{x_0}, G)/Ad \ \rightarrow \ K_{\cal L}/Ad \cr
 & \phi_1([h])  = [(h \mid_{S^*})_{S^* \in \cal L}]  \cr} \eqno(3.12)
$$

\vskip 0.5cm

\vbox{

\item{} {\bf Lemma 3.3}

\item{} The map
$$
\eqalign{ & \phi_2 \ : \  K_{\cal L}/Ad     \ \rightarrow \ H_{\cal L} \cr
 & \phi_2\left([(h_{S^*})_{S^* \in \cal L}]\right) \cr &
 = ([h_{S^*}])_{S^* \in \cal L}  \cr} \eqno(3.13)
$$
is bijective.}

\vskip 0.5cm

\vbox{

\item{} {\bf Proof of Lemma 3.3}

\item{} We will first show that $\phi_2$ is surjective.
To do so, recall that  any element
of $H_{\cal L}$ is a family $([h_{S^*}])_{S^*}$ of consistent
equivalence classes in the sense of (3.7b).  Now, choose a representative
$h^0_{S^*}$ from each $[h_{S^*}]$ and construct the subgroup $C_{S^*}^0$
of $G$ that commutes with $h^0_{S^*}$; that is, let}
$$
C^0_{S^*} = \{g \in G : \ \forall \alpha \in S^* , \ gh^0_{S^*}(\alpha)g^{-1}
= h^0_{S^*}(\alpha)\} \eqno(3.14)
$$
\item{} Note that $C^0_{S^*}$ is closed in $G$.
Any closed subgroup of a Lie group is
a Lie group and any closed subset of a compact space is compact, so that
$C^0_{S^*}$ is again a compact Lie group.  Thus, $C^0_{S^*}$
has some dimension $d_{S^*} \geq 0$ and, by compactness,
some finite number $m_{S^*} \geq 1$ of connected components.
There is then some least value $d_0$ of $d_{S^*}$
($d_0 = \min_{S^* \in \cal L} d_{S^*}$)
and some $m_0$
that is the least
value of $m_{S^*}$ for which the dimension of $C^0_{S^*}$ is $d_0$
(i.e.
$m_0 = \min_{d_{S^*} = d_0 } m_{S^*}$).
Choose some $S^*_0$ with $d_{S^*_0} = d_0$ and $m_{S^*_0} = m_0$.

\item{}Now, for every $S^* \supset S^*_0$, choose another representative
$h^1_{S^*}$ of $[h_{S^*}]$ such that
$$
h^1_{S^*} \mid_{S^*_0} = h^0_{S^*_0} \eqno(3.15)
$$
and construct the corresponding $C^1_{S^*}$:
$$
C^1_{S^*} = \{ g \in G : \ \forall \alpha \in S_* \ gh^1_{S^*}(\alpha)
g^{-1} = h^1_{S^*}(\alpha)  \}  \eqno(3.16)
$$

\item{} Note that $C^1_{S^*} \subset C^0_{S^*_0}$ and that $C^1_{S^*}$
differs from $C^0_{S^*}$ only by conjugation.  Thus,
$C^1_{S^*}$ has dimension $d_{S^*} \geq d_{S^*_0}$ and
$m_{S^*}$ connected components.  But, since
$C^1_{S^*}$ is contained in $C^0_{S^*_0}$, $d_{S^*_0} \geq d_{S^*}$
so that $C^1_{S^*}$ and $C^0_{S^*_0}$ are of the same dimension.
It follows that they agree in some neighborhood of the identity and thus on
the entire component connected to the identity.
Since $C^0_{S^*_0} \supset C^1_{S^*}$ is a disjoint union of $m_{S^*_0}$
copies of this component,
$m_{S^*} \leq m_{S^*_0}$.  But, since $C^1_{S^*}$ has dimension $d_0$,
we have $m_{S^*} \geq m_{S^*_0}$ and in fact $m_{S^*} = m_{S^*_0}$.  We thus
conclude that $C^1_{S^*} = C^0_{S^*_0}$.

\item{} This means that $h^1_{S^*}$ is unique, since any $g$ that
commutes with $h^0_{S^*_0}(\alpha) = h^1_{S^*}(\alpha) $
for all $\alpha \in S^*_0$ lies in $C^0_{S^*_0} = C^1_{S^*}$ and commutes
with $h^1_{S^*}(\alpha)$ for all $\alpha \in S^*$.  Thus, no other
representative of $[h_{S^*}]$ satisfies (3.15).  It now
follows that for any $S^*{}' \supset S^* \supset S^*_0$,
$$
h^1_{S^*{}'} |_{S^*} = h^1_{S^*} \eqno(3.17)
$$
since  $h^1_{S^*{}'}|_{S^*}$ is the unique representative of
$[h^1_{S^*}]$ that satisfies
$$
(h^1_{S^*{}'}\mid_{S^*}) \big|_{S^*_0} = h^1_{S^*{}'}|_{S^*_0}
= h^0_{S^*_0}. \eqno(3.18)
$$

\item{} Finally, for any $S^*$ that does not contain $S^*_0$, let $S^*{}'$
be any subgroup of ${\cal HG}_{x_0}$ generated by a finite number of
independent hoops that contains $S^*$
and $S^*_0$ (we see from [6] that such a group exists) and let
$$
h^1_{S^*} = h^1_{S^*{}'}|_{S^*}. \eqno(3.19).
$$
Then the
representatives $(h^1_{S^*})_{S^* \in \cal L}
 \in ([h_{S^*}])_{S^* \in \cal L}$
 form a consistent family of homomorphisms
in $K^{({\cal L})}$ and the equivalence class of this family
under the adjoint action is a member of
$K_{\cal L}/Ad$ that maps to $([h_{S^*}])_{S^* \in \cal L}$
under the map $\phi_2$.
We conclude that $\phi_2$ is surjective.

\item{} Now, injectivity of $\phi_2$ follows in a straightforward fashion.
Consider any other equivalence class of families $[(h'_{S^*})_{S^*\in
{\cal L}}] \in K_{\cal L}/Ad$ that maps to the family
$([h_{S^*}])_{S^* \in \cal L}$ chosen above under
$\phi_2$.  As with the family constructed above,
$h^1_{S^*_0}$ must be a representative of  $[h'_{S^*_0}]$.  Let
$(h^2_{S^*})_{S^* \in \cal L}$ be any family in
$[(h'_{S^*})_{S^* \in \cal L}]$
such that $h^2_{S^*_0} = h^1_{S^*_0}$.  We have just
seen that $(h^1_{S^*_0})_{S^* \in \cal L}$
is the unique self-consistent family
of homomorphisms that includes $h^1_{S^*}$ and satisfies
$[h^1_{S^*}] = [h_{S^*}]$.  Therefore, $h^2_{S^*} = h^1_{S^*}$
and the families $[(h^2_{S^*})_{S^* \in \cal L}]$ and
$[(h^1_{S^*})_{S^* \in \cal L}]$
coincide, showing that $\phi_2$ is also injective.
$$
\hskip 7cm
Q.E.D.
$$

\vskip 0.5cm

\vbox{

\item{} We complete the proof of the proposition by
noticing that the
bijective map $\phi$ is given by
$$
\phi = \phi_2 \circ \phi_1   \eqno(3.20)
$$
$$
\hskip 7cm        Q.E.D.
$$}

\vskip 1cm

Endowed with the natural topology,
the spaces $H_{S^*}$ are compact topological spaces (see (3.6)).
The Tychonov topology $\tau_T$ on the product space $H^{({\cal L})}$
is the minimal topology for which all the projections
$$
\eqalign{
\pi_{S^*} \ &: \ H_{{\cal L}} = \overline {{\cal A}/{\cal G}}
\rightarrow H_{S^*}  \cr
\pi_{S^*} ([h]) & = [h \mid_{S^*}] \cr} \eqno(3.21)
$$
are continuous. It coincides with the topology of
pointwise convergence in $H^{({\cal L})}$, i.e  the net
$[h]^{(\nu)} = ([h_{S^*}]^{(\nu)})_{S^* \in {\cal L}}$
is $\tau_T$-convergent
$$
\eqalignno{
[h]^{(\nu)} \ &{\buildrel \tau_T \over \rightarrow} \ [h]  \cr
\noalign{\hbox{if and only if}}
[h_{S^*}]^{(\nu)} \ & \rightarrow \ [h_{S^*}] \ , \forall S^* \in {\cal L}
\ , & (3.22) \cr}
$$
where the last convergence is with respect to the topology on
$H_{S^*} = G^n / Ad$. In this topology, the space
$H^{({\cal L})}$ is compact
(see [8, Tychonov theorem]). Let us also refer to
the topology induced on the projective limit $H_{\cal L} \subset
H^{({\cal L})}$
from $H^{({\cal L})}$ as the Tychonov topology $\tau_T$.
Then from the continuity of the projections
$p_{S^*S^{*'}}$, $H_{\cal L}$ is closed in $H^{({\cal L})}$ and
therefore
$$
(H_{\cal L} , \tau_T) \eqno(3.23)
$$
is also a compact topological space.
Since $H_{\cal L}$ is compact in the Tychonov topology
and
$\overline {{\cal A} /
{\cal G}}$ is compact
 in the Gel'fand topology $\tau_{Gd}$ it is natural to expect
that the bijective map $\phi$ in (3.9) is actually a homeomorphism.
Indeed we have

\bigskip

\vbox{

\item{} {\bf Proposition 3.4}
\item{} The bijective map in (3.9) is a homeomorphism
$$
\phi \ : \ ( \overline {{\cal A} / {\cal G}} \ , \tau_{Gd} ) \rightarrow
(H_{\cal L} , \tau_T) \ , \eqno(3.24)
$$
where $\tau_{Gd}$ and $\tau_T$ denote the Gel'fand and Tychonov topologies
respectively.}

\vskip 0.5cm

\vbox{

\item{} {\bf Proof}

\item{} First let us obtain a more convenient characterization
of the topology on the spaces $H_{S^*}$. As mentioned
above, $H_{S^*}$ endowed with the standard topology induced from
$G^n$ is a compact Hausdorff space. Consider  on $H_{S^*}$
the continuous functions
$$
T^{S^*}_\a([h_{S^*}]) = Tr (h_{S^*}(\a))  \ , \qquad \a \in S^* \  .
\eqno(3.25)$$
They separate the points in  $H_{S^*}$ for the same reason that
the $T_\a \ , \ \a \in {\cal HG}_{x_0}$, separate the points in
$ \overline {{\cal A} / {\cal G}}$ [2, 6]. Therefore, according to the
Stone-Weierstrass theorem [9] the algebra ${\cal HA}_{S^*}$ obtained
by taking finite linear combinations (with complex coefficients)
and products of $T_\a^{S^*}$ is dense in the $C^*$-algebra
$C(H_{S^*})$ of all continuous functions on  $H_{S^*}$ i.e.
$$
\widetilde {\cal HA}_{S^*} = C(H_{S^*}) \ . \eqno(3.26)
$$
Using the first Gel'fand-Naimark theorem [2, 9, 10] we then conclude
that the spectrum of $\widetilde {\cal HA}_{S^*}$, endowed
with the Gel'fand topology (see below) is homeomorphic
to $H_{S^*}$. An equivalent description of
the
initial topology in $H_{S^*}$ is therefore given by the Gel'fand topology,
which is, by definition, the weakest for which all
the functions $T^{S^*}_\a \ , \ \a \in S^*$ are continuous.

\item{} Returning to (3.24) we see that, in accordance with (3.21), the
Tychonov
topology on $H_{\cal L}$ is the weakest for which all the functions
$T_\a^{S^*} \circ \pi_{S^*} \ : \ H_{\cal L} \rightarrow \C$
$\ \a \in S^*, S^* \in {\cal L}$
are continuous. On the other hand the Gel'fand topology
on $ \overline {{\cal A} / {\cal G}}$ is the weakest for
which all the functions $T_\a \ , \a \in {\cal HG}_{x_0}$ are
continuous.
Since for all $\a \in {\cal HG}_{x_0}$
$$
T_\a \circ \phi^{-1} = T_\a^{S^*} \circ \pi_{S^*} \ , \forall S^* \ :
\a \in S^*. \eqno(3.27)
$$
we conclude that $\phi$ in (3.9) is a homeomorphism.
$$
\hskip 7cm
Q.E.D.
$$}

\vskip 1cm

We now proceed to derive a measure theoretic analog of proposition 3.4.
Let ${\cal B}_{S^*}$ denote the Borel $\sigma$-algebra
on $H_{S^*}$ so that,
since the projections $p_{S^*S^{*'}}$ are measurable,
$$
\lbrace (H_{S^*}, {\cal B}_{S^*}) , p_{S^*S^{*'}} \rbrace_{S^*S^{*'} \in
\cal L}      \eqno(3.28)
$$
is a projective family of $\sigma$-measurable spaces (see (2.19)). Let
$$
(H_{\cal L} , {\cal B}_{\cal L}) \eqno(3.29)
$$
denote the projective limit $\sigma$-measurable space. In
$\overline {{\cal A} /
{\cal G}}$ we take the measurable sets to be
generated by the class ${\overline {\cal C}}$ of
``cylindrical sets" used in [6, 7],
i.e. the inverse images $C_B$ of Borel sets $B$ in $G^n/Ad$ with respect to
$\pi_{S^*} \circ \phi$
$$
\eqalignno{
C_B & \in \overline {\cal C} \quad \Leftrightarrow & (3.30a) \cr
 C_B  = (\pi_{S^*} \circ \phi )^{-1} (B) &=
\lbrace [h] \in \overline {{\cal A} /
{\cal G}} \ : \ [h(\beta_1) , \ldots , h(\beta_n)] \in B \subset G^n / Ad
\rbrace    \ , & (3.30b) \cr}
$$
where, as in (3.5),  we have identified $H_{S^*}$ with $G^n / Ad$
 with the help of the independent hoops
$$
\beta_1 , \ldots , \beta_n \in S^* \ .
$$
Note that the complement of a cylindrical set is cylindrical, as are
finite unions and intersections of cylindrical sets so that
${\overline {\cal C}}$ is in fact a (finite) algebra.
Denoting the minimal $\sigma$-algebra
algebra containing the cylindrical
sets by ${\cal B}(\overline {\cal C})$, the space
$$
(\overline {{\cal A} /
{\cal G}} , {\cal B}(\overline {\cal C}))   \eqno(3.31)
$$
becomes a $\sigma$-measurable space.  From the definition
of ${\cal B}_{\cal L}$ and
${\cal B}({\overline {\cal C}})$, we see  that

\bigskip

\vbox{

\item{}{\bf Proposition 3.5}

\item{} The map (3.9)
$$
(\overline {{\cal A} / {\cal G}}, {\cal B}(\overline {\cal C})) \rightarrow
(H_{\cal L} , {\cal B}_{\cal L}) \eqno(3.32)
$$
is an isomorphism of $\sigma$-measurable spaces.}

\vskip 0.5cm

\vbox{

\item{} {\bf Corollary 3.6
}

\item{(i)} ${\cal B}(\overline {\cal C})$ and ${\cal B}_{\cal L}$
are contained in  the Borel algebras corresponding to the Gel'fand and
Tychonov topologies respectively. This follows from the fact that the
cylindrical sets in ${\cal B}_{\cal L}$ with open ``base"
$B$ in ${\cal B}_{S^*}$ form a base in the topology
$\tau_T$.

\item{(ii)} We call a function $f$ on
$\overline {{\cal A} / {\cal G}}$ cylindrical if there
exists $S^* \in \cal L$ such that $f$ is a pull back of
a function $\tilde f$ on $H_{S^*}$
$$
\eqalignno{
f & = (\pi_{S^*} \circ \phi )^* \tilde f    & (3.33a) \cr
\noalign{\hbox{\qquad i.e.}}
f([h]) & = \tilde f ([h \mid_{S^*}]) \  ,    & (3.33b) \cr}
$$
where $\tilde f$ is a measurable function on
$H_{S^*}$. The Wilson loop functions \break
$T_\alpha([h]) = {1 \over N} Tr h(\a)$
(for $G = SU(N)$ or $G = U(N)$)
are continuous cylindrical
functions ([6]).

\item{(iii)} The projective limit $H_{\cal L}$ provides a generalization of
the Ashtekar-Isham space $\overline {{\cal A}/{\cal G}}$ to the case
where the gauge group $G$ is not compact.

\item{(iv)} There is a one-to-one correspondence between cylindrical
measures $\mu$ on $\overline {\cal C}$ (i.e. additive
on $\overline {\cal C}$ but $\sigma$-additive on the $\sigma$-subalgebras
$(\pi_{S^*} \circ \phi)^{-1}({\cal B}_{S^*})$)
and families of measures
$\{ (\mu_{S^*})_{S^* \in \cal L} \}$ ($\mu_{S^*}$ are Borel measures
on the {\it finite dimensional} spaces $H_{S^*}$) satisfying
the self-consistency condition}
$$
S^* \subset S^{*'} \ \Rightarrow \ \mu_{S^*} = (p_{S^*S^{*'}})_*
\mu_{S^{*'}}     \ . \eqno(3.34)
$$
\item{} The correspondence is given by
$$
\mu_{S^*} = (\pi_{S^*} \circ \phi)_*  \mu  \ .     \eqno(3.35)
$$

\vskip 1cm

Recall [11] that a Borel measure $\mu$ is called regular
if for every Borel set $E$
$$
\eqalignno{
\mu(E) & = inf \{ \mu(V) \ : \ E \subset V \ , V \ open \} \cr
\mu(E) & = sup \{ \mu(K) \ : \ E \supset K \ , K \ compact \} \ . \cr}
$$
Also from [11, Theorem 2.18] it follows that on the
spaces $H_{S^*}$ every Borel measure is regular.
The following result (similar to [6, Theorem 4.4] and
[7, Proposition 2]) holds.

\vbox{

\vskip 0.5cm

{\bf Proposition 3.7}

\item{} There is a one-to-one correspondence    between
regular Borel
measures $\mu$ on $\agb$ and self-consistent families of measures
$\{(\mu_{S^*})_{S^* \in \cal L}\}$.}

\vskip 0.5cm

\vbox{

{\bf Proof}

\item{} From (i) and (iv) we see that a regular Borel measure $\mu$
on $\agb$ defines, by restriction, a $\sigma$-additive measure
on ${\cal B}(\overline {\cal C})$ and therefore a consistent family
of measures $\{(\mu_{S^*})_{S^* \in \cal L} \} $. Conversely let
$\{(\mu_{S^*})_{S^* \in \cal L} \}$ be a consistent family of Borel measures
on $\{ H_{S^*} \}$ and $\mu_0$ be the cylindrical measure on $\overline
{\cal C}$
defined by this family. The family $\{(\mu_{S^*})_{S^* \in \cal L} \}$
(or equivalently the measure $\mu_0$) defines a positive functional
on the continuous cylindrical functions $f = (\pi_{S^*} \circ \phi)^*
\tilde f$ on $\agb$ }
$$
\Gamma_{\mu_0} (f) = \int_{H_{S^*}} \tilde f d \mu_{S^*} \ . \eqno(3.36)
$$
\item{} This functional is bounded with respect to the sup-norm
$$
\mid \Gamma_{\mu_0}(f) \mid \leq \parallel f \parallel_\infty  \ ,
\eqno(3.37)
$$
where $\parallel f \parallel_\infty = \sup_{[h] \in \agb} \mid f([h]) \mid$.
Since the space of continuous cylindrical functions is
dense  in the  $C^*$-algebra
$C(\agb)$ of all continuous functions on $\agb$ (see [6])  the functional
$\Gamma_{\mu_0}$ can be extended in a unique way to
a continuous positive (norm $1$) functional on $C(\agb)$ (see [9]). But
in accordance with the Riesz representation theorem (see [11])
there is then  a unique regular Borel measure $\mu$ on
$\agb$ such that
$$
\Gamma_{\mu_0}(f) = \int_{\agb} d \mu f   \eqno(3.38)
$$
for every $f \in C(\agb)$, where we denoted the extension
of $\Gamma_{\mu_0}$ to $C(\agb)$ with the same letter.
Regular Borel measures are completely determined
if the integral of continuous functions is known (see [11], p.41),
which implies that $\mu$ and $\mu_0$ coincide on
$\overline {\cal C}$. Therefore $\mu$  is the unique (see [12]) extension of
$\mu_0$ to ${\cal B}({\cal C})$ and (as we have showed) the unique regular
extension to a Borel  measure.
$$
\hskip 7cm Q.E.D.
$$

\vskip 1.5cm

\vbox{

{\bf 4.  $ {{\cal A} / {\cal G}}$ is contained in a zero measure subset of
$\overline {{\cal A} / {\cal G}}$.}

\vskip 0.5cm

The present section contains the main result of this paper. For simplicity
we will use (3.32) to identify the $\sigma$-measurable spaces
$(\overline {{\cal A} / {\cal G}}, {\cal B}(\overline {\cal C}))$ and
$(H_{\cal L}, {\cal B}_{\cal L})$ so that we will consider
$\overline {{\cal A} / {\cal G}}$ to be the projective limit of the
projective family of finite dimensional spaces (3.6).}

In [6] Ashtekar and Lewandowski introduced the following measure
$\mu_{AL}$ on \break
$(\overline {{\cal A} / {\cal G}}, {\cal B}(\overline {\cal
C}))$. Let $\mu_H$ be the  normalized Haar measure  on $G$ and $\mu^H_n
$ and $ \mu^H_{S^*}$
the corresponding measures on $G^n / Ad$  and
$H_{S^*}$   ($\mu^H_{S^*}$ is obtained from $\mu^H_n$ using (3.5)).
Then the (uncountable) family $(\mu^H_{S^*})_{S^* \in \cal L}$ satisfies
the self-consistency conditions (2.20). The Ashtekar-Lewandowski measure
$\mu_{AL}$ is the corresponding (unique) measure
on $(\overline {{\cal A} / {\cal G}}, {\cal B}(\overline {\cal C}))$
satisfying
$$
\mu^H_{S^*} = (\pi_{S^*})_*  \mu_{AL}      \ .  \eqno(4.1)
$$
The measure $\mu_{AL}$ is $\sigma$-additive, $Diff(\Sigma)$-invariant, and
strictly positive as a functional on the
space continuous cylindrical functions on $\overline {{\cal A}/{\cal G}}$
(see [6]).

The space $ {{\cal A} / {\cal G}}$ is canonically embedded in
$\overline {{\cal A} / {\cal G}}$ [2] and is topologically dense there
[6, 10]. It is interesting to find out whether ${{\cal A} / {\cal G}}$
is also $\mu_{AL}$-thick in $\overline {{\cal A} / {\cal G}}$; that is,
whether ${\cal A}/{\cal G}$ supports the measure $\mu_{AL}$.
We will in fact prove that this is far from being the case:

\vskip 0.5cm

\vbox{

\item{} {\bf Theorem 4.1}

\item{} There exists a measurable set
$$
\eqalignno{
Z &\in {\cal B}(\overline {\cal C}) & (4.2a) \cr
\noalign{\hbox{ \qquad such that}}
\mu_{AL}(Z) &= 0   & (4.2b) \cr
\noalign{\hbox{ \qquad and}}
{{\cal A} / {\cal G}}    &\subset Z    \ .   & (4.2c) \cr}
$$
}

\vskip 0.5cm

\vbox{

\item{} {\bf Proof}

\item{} We need the following lemma}

\vskip 0.5cm

\vbox{

\item{} {\bf Lemma 4.2}

\item{} For every $q \in (0, 1]$ there exists $Q^{(q)} \subset
\overline {{\cal A} / {\cal G}}$  such that
$$
\eqalignno{
\mu_{AL} (Q^{(q)}) & = q         & (4.3)   \cr
\noalign{\hbox{\qquad and}}
 {{\cal A} / {\cal G}} & \subset Q^{(q)}     & (4.4) \cr}
$$
}
\vskip 0.5cm

\vbox{

\item{} {\bf Proof of Lemma}

\item{} The complement $Q^{(q)^c}$ of $Q^{(q)}$ will be constructed
essentially (i.e. modulo dividing by $Ad$) by taking an infinite
product of sets consisting of copies of $G$ with holes cut out
around the identity such that the ``diameter" of the holes
decreases to zero. These copies of $G$ are chosen to correspond
to a certain ``convergent" sequence of hoops.
In order to do this explicitly,
choose $r_0$ such that
the exponential map is one-to-one
in the subset $\overline {{\cal U}_{r_0}(0)}$
of $ Lie(G)$
where}
$$
\eqalignno{
\overline {{\cal U}_{r_0}(0)} & = \lbrace  v \in \ Lie (G)  : \
\parallel v \parallel \leq r_0 \rbrace     \cr
\noalign{   \hbox{ \qquad and}}
exp \ & : \ \overline {{\cal U}_{r_0}(0)} \rightarrow
\overline {{\cal O}_{r_0}(e)}   \subset G  \ ,  & (4.5) \cr}
$$
\item{} that is, ${\overline {\cal O}_{r_0}(e)}$ is the image of
${\overline {\cal U}_{r_0}(0)}$ under the exponential map,
where $e$ is the identity of the group.
Here $r_0 > 0$ and $\parallel \cdotp \parallel $ denotes the
norm induced by a bi-invariant inner product in $Lie (G)$
(the Killing form if $G$ is semisimple).

\item{} Let us define a function on
$\overline {{\cal O}_{r_0}(e)}$ that measures
the ``distance" to the identity $e$
$$
\eqalign{
d_e \ & : \ \overline {{\cal O}_{r_0}(e)} \rightarrow \R^+ \cup \{0\}  \cr
d_e(g) & = \parallel ln (g) \parallel \cr} \eqno(4.6)
$$
and denote by the same letter $d_e$ the following extension
to the whole group $G$:
$$
\eqalignno{
d_e \ & : \ G \rightarrow \R^+ \cup \{0\}   & (4.7a) \cr
d_e(g) & = \Bigg\{ {r_0 \qquad  g \in {{\cal O}_{r_0}(e)^c}
\atop \parallel ln(g) \parallel \quad g \in {{\cal O}_{r_0}(e)} }    &
(4.7b) \cr}
$$
The $Ad$-invariance of $\parallel \cdotp \parallel $ on $Lie (G)$
implies that $d_e( \cdotp )$ is $Ad$-invariant on $G$. Consider now
the basic sets
$$
\eqalign{
\Delta^\epsilon & \subset G   \cr
\Delta^\epsilon & = \lbrace g \in G \ : \ d_e(g) \geq \epsilon
\rbrace \quad 0 \leq \e \leq r_0  . \cr} \eqno(4.8)
$$
The function given by
$$
\eqalign{
s & \ : \ [0, r_0 ) \rightarrow \R^+ \cr
s(\epsilon ) & = \mu_H ( \Delta^\epsilon ) \cr} \eqno(4.9)
$$
is continuous, monotonically decreasing and
$s(0) = 1$.
Now let
$\Delta^{\{ \epsilon_i \}^n_{i=1}}_n $ be the subset of $G^n$ given by
$$
\eqalign{
\Delta^{\{ \epsilon_i \}^n_{i=1}}_n & = \lbrace (g_1, \ldots ,
g_n) \ : \ d_e(g_i) \geq \epsilon_i \rbrace
= \prod_{i=1}^n \Delta^{\epsilon_i} \  .  \cr} \eqno(4.10)
$$
Clearly we have
$$
\mu^H_n(\Delta^{\{ \epsilon_i \}^n_{i=1}}_n) = \prod_{i=1}^n s(\epsilon_i)
\ . \eqno(4.11)
$$
Notice
that the set $\Delta^{\{ \epsilon_i \}}_n$ is an
$Ad$-invariant subset of $G^n$. It is the inverse image of the set
$$
\eqalignno{
\widetilde \Delta^{\{ \epsilon_i \}^n_{i=1}}_n & \subset G^n/Ad
 & (4.12a) \cr
\widetilde \Delta^{\{ \epsilon_i \}^n_{i=1}}_n & = \lbrace [g_1, \ldots ,
g_n] \ : \ d_e(g_i) \geq \epsilon_i \rbrace  \  & (4.12b) \cr}
$$
under the quotient map $\pi : G^n \rightarrow G^n/Ad$.
By the definition of the measure $\mu^H_n$ on $G^n/Ad$ we thus have
$$
\mu^H_n (\tilde \Delta^{\{ \epsilon_i \}_{i=1}^n}_n) = \prod_{i=1}^n s
(\epsilon_i) \ . \eqno(4.13)
$$
Now, for each $q \in (0, 1]$ choose a sequence
$$
\eqalignno{
\lbrace \epsilon^{(q)}_i \rbrace_{i=1}^\infty&    & (4.14a) \cr
\noalign{\hbox{\qquad such that \ $ \epsilon_i^{(q)} \neq 0$ but}}
\lim_{i \rightarrow \infty} \epsilon^{(q)}_i & = 0 & (4.14b) \cr
\noalign{ \hbox{\qquad and}}
1-q & = \lim_{n \rightarrow \infty} \prod_{i=1}^n s(\epsilon^{(q)}_i)
& (4.14c) \cr}
$$
Let $\{\b_i\}_{i=1}^{\infty}$ be an arbitrary sequence
of independent hoops. Then
the sets
$$
\eqalign{
\widehat \Delta^{\{\epsilon^{(q)}_i\}^n_{i=1}}_n & \ \subset
\ \overline {{\cal A} / {\cal G}}    \cr
\widehat \Delta^{ \{ \epsilon^{(q)}_i\}^n_{i=1}}_n & =
(\pi_{S^* [ \beta_1, \ldots , \beta_n ] } \circ \phi)^{-1}
\left( \widetilde \Delta^{ \{ \epsilon^{(q)}_i\}^n_{i=1}}_n  \right)
 \ , \cr} \eqno(4.15a)
$$
were we used (3.5) to identify $H_{S^*}$ and $G^n/Ad$,
form a decreasing sequence
$$
\widehat \Delta^{ \{ \epsilon^{(q)}_1\} }_1 \supset \ldots \supset
\widehat \Delta^{ \{ \epsilon^{(q)}_i\}^n_{i=1}}_n \supset \ldots
\eqno(4.15b)
$$
such that
$$
\mu_{AL} \left(
\widehat \Delta^{ \{ \epsilon^{(q)}_i\}^n_{i=1}}_n \right) =
\mu_n^H \left(
\widetilde \Delta^{ \{ \epsilon^{(q)}_i\}^n_{i=1}}_n \right)
\ . \eqno(4.15c)
$$
Now, introducing $R^{(q)}(\{\beta_i\})$ whose complement in $\agb$ is
$$
R^{(q)}(\{\b_i\})^c = \cap_{n=1}^\infty \
\widehat \Delta^{ \{ \epsilon^{(q)}_i\}^n_{i=1}}_n   \ , \eqno(4.16)
$$
we conclude from (4.13) and the $\sigma$-additivity
of $\mu_{AL}$ that
$$
\eqalignno{
\mu_{AL}(R^{(q)}(\{\b_i\})^c) &= \lim_{n \rightarrow \infty} \mu_{AL}(
\hat{\Delta}_n^{\{ \epsilon_i^{(q)}\}_{i=1}^n }) = 1-q   \cr
\noalign{\hbox{\qquad and}}
\mu_{AL} (R^{(q)}(\{\beta_i\})) & = q \ \in \ (0,1] \ . & (4.17) \cr}
$$
Let us now turn to the second part of the lemma namely the choice
of $Q^{(q)}$ satisfying (4.3) and (4.4). Take for $\hat \beta_i$
the hoops corresponding to
coordinate squares (all parallel to a fixed coordinate plane) with a corner
at $x_0$ and fix  a metric. Choose $\hat \beta_i$ to have areas such that
$$
Area(\hat \beta_i) = \epsilon^{(q)}_i \delta_i \ , \eqno(4.18)
$$
where
$\{ \epsilon^{(q)}_i \}^\infty_{i=1}$ is the same as in (4.14)
and $\{ \delta_i \}^\infty_{i=1}$ is any
sequence with $\delta_i \rightarrow 0$.
Let
$$
Q^{(q)} = R^{(q)}(\{\hat \b_i\})   \ .
$$
Then,  for every $A \in {\cal A}$ we have (from the smoothness
of $A$)
$$
H(\hat \beta_i, A) = 1 + F(A) \epsilon_i^{(q)} \delta_i +
O(\epsilon^{(q)^2}_i \delta^2_i)       \ , \eqno(4.19)
$$
where $F(A)$ denotes the component of the curvature at $x_0$
in the plane of the squares $\hat \beta_i$. Then for every
$[A] \in  {{\cal A} / {\cal G}}$ there exists a constant $c([A]) > 0$
such that
$$
d_e (H(\hat \beta_i, [A])) < c([A]) \epsilon^{(q)}_i \delta_i    \ .
\eqno(4.20)
$$
and, since $\delta_n \rightarrow 0$, for $n$ large enough  we have
$$
d_e(H(\hat \beta_n, [A])) < \epsilon^{(q)}_n \ .
$$
Thus, for every $[A] \in {\cal A}/{\cal G}$, $[A] \in Q^{(q)}$.

We have therefore proved that with our choice (4.18)
of $\hat \beta_i$
we have
$$
 {{\cal A} / {\cal G}} \subset Q^{(q)}   \eqno(4.4)
$$
$$
\hskip 7cm Q.E.D.
$$

\vskip 0.5cm

\item{} Let us now prove the theorem. From (4.4) we conclude that
for every $q > 0$
$$
\eqalignno{
 {{\cal A} / {\cal G}} & \subset Q^{(q)} \subset
\overline {{\cal A} / {\cal G}}      & (4.21a) \cr
\noalign{\hbox{\qquad and}}
\mu_{AL}(Q^{(q)}) &= q    \ . & (4.21b) \cr}
$$
Considering now the decreasing sequence $Q^{(1/n)}$.  We have
$$
 {{\cal A} / {\cal G}} \subset Z \equiv \cap_{N=1}^\infty \ Q^{(1/n)} \ .
 \eqno(4.22)
$$
while the $\sigma$-additivity of $\mu_{AL}$ implies that
$$
\mu_{AL} (Z) = \lim_{N \rightarrow \infty}
 \mu_{AL}(Q^{(1/n)}) = 0
\eqno(4.23)
$$
$$
\hskip 7cm Q.E.D.
$$

\vbox{

{\bf 5. Completion of the space  of
square integrable functions on $ {{\cal A} / {\cal G}}$}

\vskip 0.5cm

Although ${{\cal A} / {\cal G}} $ is not a projective limit
of the family (3.6) a procedure similar to that of (2.14),
(2.19)-(2.21) can be used to define a measure $\hat \mu_{AL}$
on ${{\cal A} / {\cal G}} $ as was noted in [6]. This is done
by returning to the notion of a cylindrical set (3.17) but
now in ${{\cal A} / {\cal G}}$. That is, we introduce
(surjective)  projections}
$$
\eqalign{
\hat \pi_{S^*} \ : \ {{\cal A} / {\cal G}} & \rightarrow
H_{S^*[\beta_1 , \ldots , \beta_n ]}    \cr
\pi_{S^*}([A]) & = [H(\beta_1, A), \ldots , H(\beta_n, A)] \ ,  \cr}
\eqno(5.1) $$
where again we are identifying $G^n/Ad$ with $H_{S^*}$,
and take as measurable sets
$$
\eqalign{
C_B & \subset {{\cal A} / {\cal G}} \cr
C_B & = \hat \pi_{S^*}^{-1} (B)  \ , \cr} \eqno(5.2)
$$
for some $B \in {\cal B}_{S^*}$. Let ${\cal C}$ be the collection
of such cylindrical sets in ${{\cal A} / {\cal G}}$. Note
that ${\cal C}$ is closed under union, intersection and complementation
(i.e. forms an algebra) so that the pair
$$
({{\cal A} / {\cal G}},
{\cal C})    \eqno(5.3)
$$
is a measurable space. The measure $\hat \mu_{AL}$ is then defined by
$$
\hat \mu_{AL} (\hat \pi_{S^*}^{-1}(B)) = \mu_{S^*}^H (B)  \ . \eqno(5.4)
$$
The additivity of $\mu_{S^*}^{H}$ for every $S^*$ implies
additivity of $\hat \mu_{AL}$. However, the $\sigma$-additivity of
the $\mu_{S^*}^H$ does not imply $\sigma$-additivity
of $\hat \mu_{AL}$. Indeed, we have the following

\vskip 0.5cm

\vbox{

{\bf Proposition 5.1}

\item{} The measure (5.4) on ${{\cal A} / {\cal G}}$
cannot be extended to a $\sigma$-additive measure on
${\cal B} ({\cal C})$.}

\vskip 0.5cm

\vbox{

{\bf Proof}

\item{} This theorem follows easily from lemma 4.2. Indeed consider the same
sets
$\widetilde \Delta^{\{ \epsilon_i^{(q)} \}^n_{i=1}}_n \subset G^n /Ad$
as in (4.12)-(4.14) and define analogously to (4.15)
the decreasing sequence}
$$
\eqalignno{
\dot  \Delta^{\{ \epsilon_i^{(q)} \}^n_{i=1}}_n \ &\subset
{{\cal A} / {\cal G}} & (5.5a)   \cr
\dot \Delta^{\{ \epsilon_i^{(q)} \}^n_{i=1}}_n
& = \hat \pi^{-1}_{S^*[\hat \beta_1 ,  \ldots , \hat \beta_n ]} \left(
\widetilde  \Delta^{\{ \epsilon_i^{(q)} \}^n_{i=1}}_n \right)
& (5.5b)   \cr
\dot  \Delta^{\{ \epsilon_1^{(q)} \}}_1  & \supset
\ldots  \supset
\dot  \Delta^{\{ \epsilon_i^{(q)} \}^n_{i=1}}_n
\supset  \ldots \ , & (5.5c)  \cr}
$$
\item{} where the sequence $\{ \hat \beta_i \}_{i=1}^\infty $ is defined
as in (4.18). Then for the same reason as in (4.20)
there is not a single $[A]$ belonging to
the intersection of all $\dot  \Delta^{\{ \epsilon_i^{(q)} \}^n_{i=1}}_n$
i.e. now we have
$$
\cap_{n=1}^\infty \dot  \Delta^{\{ \epsilon_i^{(q)} \}^n_{i=1}}_n
= \emptyset
\eqno(5.6)
$$
even though
$$
\lim_{n \rightarrow \infty} \hat \mu_{AL}\left(
\dot  \Delta^{\{ \epsilon_i^{(q)} \}^n_{i=1}}_n \right) =  1 - q
\ . \eqno(5.7)
$$
Therefore, choosing $q \ : \ 0 < q < 1$ we conclude from the Hopf theorem 2.1
that $\hat \mu_{AL} $ is not extendible to a $\sigma$-additive
measure on ${\cal B}({\cal C})$.
$$
\hskip 7cm Q.E.D.
$$

\vskip 1cm

Let us recall aspects of integration theory
for the so called (non-$\sigma$) measurable spaces
with limit structure (see [13] def. 1.5).
The measurable space $(X, {\cal F}_X)$ is said to be a space
with limit structure if
$$
{\cal F}_X = \cup_{L \in \cal L} {\cal B}_L   \ ,  \eqno(5.8)
$$
where for all $L \in \cal L$,  ${\cal B}_L$ is a $\sigma$-algebra
and for every $L_1, L_2 \in \cal L$, there exists a $L_3$
such that ${\cal B}_{L_1} \cup {\cal B}_{L_2} \subset {\cal B}_{L_3}$.
If the family $\{ {\cal B}_L \}_{L \in \cal L} $ does
not have a maximal element then ${\cal F}_X$ is not
a $\sigma$-algebra. Obviously every
projective limit defined as in
(2.19)-(2.21) is a measurable space with limit structure. The converse
is also true as we can see by taking  as projective
family of $\sigma$-measurable spaces (see [8] p. 20)
$$
\lbrace (X_L, {\cal B}_L), p_{LL'} \rbrace_{L,L' \in \cal L} =
\lbrace (X, {\cal B}_L), id \rbrace_{L \in \cal L}    \ .  \eqno(5.9)
$$
Though this makes the class of projective limit spaces
equivalent to that of measurable spaces with
limit structure the latter is more ``natural" for integration theory.

In a measurable space with
limit structure $(X, {\cal F}_X)$ the sets $F \in {\cal F}_X$ are called
cylindrical sets and the  map $f$ to a $\sigma$-measurable space
$(Y, {\cal B})$ is called cylindrical if there is
a $L \in \cal L$ such that
$$
f \ : \  (X, {\cal B}_L) \ \rightarrow \  (Y, {\cal B})
$$
is measurable. A measure $\mu$ on ${\cal F}_X$ is called a
quasi-$\sigma$-measure (quasi-measure in [13]) if
its restriction $\mu_L = \mu \mid_{{\cal B}_L}$ to every ${\cal B}_L
\subset {\cal F}_X$ is $\sigma$-additive.
The triple $\{(X, {\cal F}_X), \mu_X \}$, where
$(X, {\cal F}_X)$ is a measurable space with limit structure
and $\mu_X$ is a quasi-measure is called a quasi-measure space.
Let $L_0 \in \cal L$ be such that
the (complex-valued) cylindrical function
$$
f \ : \ (X, {\cal B}_{L_0}) \ \rightarrow \ (\C, {\cal B})
$$
where $\cal B$ denotes the $\sigma$-algebra of the
complex plane,
is measurable. Then a function $f$ on the quasi-measure space
$\lbrace (X, {\cal F}_X), \mu \rbrace$ is said to be
$\mu$-integrable if it is $\mu_{L_0}$ integrable in the
usual sense
$$
\int_X f d \mu(x) = \int_{X_{L_0}} f d \mu_{L_0}(x)    \ .  \eqno(5.10)
$$

\vskip 0.5cm

\vbox{

{\bf Definition 5.2}

\item{} The set of square-integrable cylindrical functions
on the quasi-measure space $\{(X, {\cal F}_X) , \mu\}$ will
be denoted through ${\cal C}L^2(X, {\cal F}_X, \mu)$.}

\vskip 0.5cm

It is easy to see that
${\cal C}L^2(X, {\cal F}_X, \mu)$ is a pre-Hilbert space with inner product
given by
$$
(f, g) = \int_X \overline {f(x)} g(x) d \mu (x) = \int_X
\overline {f(x)} g(x)
d \mu_{L_0}(x)  \ , \eqno(5.11)
$$
where $L_0$ is such that
both $f \ : \ (X, {\cal B}_{L_0}) \ \rightarrow \ (\C , {\cal B})$
and $g \ : \ (X, {\cal B}_{L_0}) \ \rightarrow \ (\C , {\cal B})$
are measurable.

\vskip 0.5cm

\vbox{

{\bf Proposition 5.3}

\item{} Suppose that we are given two quasi-measure spaces
$\{(X, {\cal F}_X), \mu_X\}$ and
$\{(Y, {\cal F}_Y), \mu_Y\}$,
where}
$$
 {\cal F}_X = \cup_{L \in \cal L} {\cal B}_L(X) \quad {\rm and}
 \quad {\cal F}_Y = \cup_{L \in \cal L} {\cal B}_L(Y)
$$
\item{} and that $ Y \subset X$. Let $\chi \ : \ {\cal F}_X \ \rightarrow \
{\cal F}_Y$ be an
isomorphism of set algebras
given by
 $\chi(B) = B \cap Y $ for $B \in {\cal F}_X$ and
such that the restriction to every ${\cal B}_L(X)$
is an isomorphism of $\sigma$-algebras
${\cal B}_L(X) \ : \ {\cal B}_L(X) \ \rightarrow \ {\cal B}_L(Y)$.
Assume also that $\mu_Y \circ \chi = \mu_X$.

\item{} Then if $\mu_X$ is extendible to a $\sigma$-additive measure
$\tilde \mu_X$ on ${\cal B}({\cal F}_X)$, the completion of
${\cal C}L^2\left(Y, {\cal F}_Y, \mu_Y \right) $ is
$L^2\left(X, {\cal B}({\cal F}_X), \tilde \mu_X \right)$.

\vskip 0.5cm

\vbox{

{\bf Proof}

\item{} Note that the map $\chi \ : \ {\cal F}_X \ \rightarrow
{\cal F}_Y$ induces a one-to-one correspondence
between the sets ${\cal X}_Y$ of characteristic functions
of sets in ${\cal F}_Y$
and ${\cal X}_X$  of characteristic functions
of sets in ${\cal F}_X$.
Further, since $\chi$ is an isomorphism of finite
set algebras, this correspondence extends to an isomorphism
over the linear spans of
${\cal X}_Y$  and  ${\cal X}_X$.
Finally, since $\chi$ preserves the measure of sets,
this correspondence preserves the inner product
in  these linear spaces. We need the following lemma.
}

\vskip 0.5cm

\vbox{

{\bf Lemma 5.4}

\item{(i)} The completion of ${\cal X}_Y$ (${\cal X}_X$) is equal
to the completion of
${\cal C}L^2\left(Y, {\cal F}_Y, \mu_Y \right) $
(${\cal C}L^2\left(X, {\cal F}_X, \mu_X \right) $).

\medskip

\item{(ii)} The space ${\cal X}_X$ is dense in
$L^2\left(X, {\cal B}({\cal F}_X), \tilde \mu_X \right) $.
 }

\vskip 0.5cm

\vbox{

{\bf Proof of Lemma}

\item{(i)} Obviously ${\cal X}_Y$ is a subset of
${\cal C}L^2\left(Y, {\cal F}_Y, \mu_Y \right) $. It is sufficient to show
that any $f \in {\cal C}L^2\left(Y, {\cal F}_Y, \mu_Y \right) $
can be represented as }
$$
f = \lim_{n \rightarrow \infty} \phi_n \ , \eqno(5.12)
$$

\item{} where $\phi_n \in {\cal X}_Y$ and the sequence converges in the
norm of ${\cal C}L^2\left(Y, {\cal F}_Y, \mu_Y \right)$. But for
$f \in {\cal C}L^2\left(Y, {\cal F}_Y, \mu_Y \right)$ there exists
a $L_0 \in \cal L$ such that $f $ belongs to the (complete) space
$L^2\left(Y, {\cal B}_{L_0}(Y), \mu_Y \mid_{{\cal B}_{L_0}(Y)} \right)$.
Since ${\cal X}_Y \mid_{{\cal B}_{L_0}(Y)} \subset {\cal X}_Y $
is dense
in $L^2\left(Y, {\cal B}_{L_0}(Y), \mu_Y \mid_{{\cal B}_{L_0}(Y)} \right)$
(see [12]) $f$ can be represented in the form (5.10).

\medskip

\item{(ii)} For a quasi-measure space
$\{(X, {\cal F}_X), \mu_X\}$ satisfying the conditions of
proposition 5.2 we  have
$$
{\cal X}_X \subset {\cal C}L^2\left(X, {\cal F}_X, \mu_X \right)
\subset L^2\left(X, {\cal B}({\cal F}_X), \tilde \mu_X \right) \ ,
\eqno(5.13) $$
where clearly all the inclusions are isometric.
It will be sufficient to prove that for every set $B \in {\cal B}({\cal F}_X)$
its characteristic function $\chi_B$ is in the $L^2$-closure
of ${\cal X}_X$. But this result follows easily from
Theorem 3.3 in [8].
$$
\hskip 7cm  Q.E.D.
$$

\vskip 0.5cm

\vbox{

{\bf Proof of Proposition}

\item{} We have an isometric isomorphism (i.e. one which preserves
the inner product) between the
spaces ${\cal X}_Y$ and ${\cal X}_X$, which are dense in
$\widetilde {\cal X}_Y = \widetilde{
{\cal C}L^2}\left(Y, {\cal F}_Y, \mu_Y \right)$
and
$L^2\left(X, {\cal B}({\cal F}_X), \tilde \mu_X \right)$
respectively.  The isomorphism therefore extends
to a natural isometric isomorphism}
$$
\eta \ : \ \widetilde{
{\cal C}L^2}\left(Y, {\cal F}_Y, \mu_Y \right) \ \rightarrow \
L^2\left(X, {\cal B}({\cal F}_X), \tilde \mu_X \right)
\eqno(5.14)
$$
$$
\hskip 7cm Q.E.D.
$$
\vskip 0.5cm

In the case of ${\cal A}/{\cal G}$ since the projections
$\hat \pi_{S^*}$ are surjective
we have
$$
\hat \pi_{S_1^*}^{-1} (B_1) = \hat \pi_{S_2^*}^{-1} (B_2)   \eqno(5.15)
$$
if and only if there is some $S^* \subset S^*_1 \cap S^*_2$ and some
$B \subset H_{S^*}$ such that
$$
B_1 =  \pi_{S^*S_1^*}^{-1} (B) \ , \ B_2 =
 \pi_{S^*S_2^*}^{-1} (B)
\eqno(5.16) $$
Since the same is true for the algebra $\overline {\cal C}$
of cylindrical sets in $\overline {{\cal A} / {\cal G}}$, there
is a one-to-one correspondence between ${\cal C}$ and
$\overline {\cal C}$ given by
$$
\hat \pi_{S^*}^{-1}(B) = \chi((\pi_{S^*} \circ \phi)^{-1} (B))     \ ,
\eqno(5.17) $$
where $\phi$ and $\pi_{S^*}$ have been defined in (3.9)
and (3.11) respectively. Note that the map $\chi$ is an isomorphism
of set algebras, and that it preserves measures in
the sense that
$$
\eqalignno{
\chi(\tilde B) & =  {{\cal A} / {\cal G}} \cap \tilde B    & (5.18a) \cr
\noalign{\hbox{ and that }}
  \hat \mu_{AL} \circ \chi & = \mu_{AL}\mid_{\overline {\cal C}} \ ,       &
(5.18b) \cr}
$$
so that the conditions of proposition 5.3 are satisfied for this case.
In this way, the completion of
\break ${\cal C}L^2\left({{\cal A} / {\cal G}}, {\cal C}, \hat \mu_{AL}
\right)$
is
$L^2\left(\overline {{\cal A} / {\cal G}}, {\cal B}(\overline {\cal C}),
\mu_{AL} \right)$ and we arrive at
the space
$\overline {{\cal A} / {\cal G}}$.

Let us also show
that
${\cal C}L^2 \left(  {{\cal A} / {\cal G}}, \hat \mu_{AL}, {\cal C} \right)$
(hereafter referred to as simply
${\cal C}L^2 \left(  {{\cal A} / {\cal G}} \right)$)
is not complete. To see this, consider the sets
$$
\eqalignno{
\widetilde \Delta_n \equiv
\widetilde \Delta^{\{ \epsilon_i^{(q)}
\}^n_{i=1}}_n & \subset G^n/Ad &(5.19a) \cr \noalign{\hbox{and}}
\dot \Delta_n \equiv
\dot  \Delta^{\{ \epsilon_i^{(q)} \}^n_{i=1}}_n \ &\subset
{{\cal A} / {\cal G}}  &(5.19b)  \cr}
$$
introduced above, for some $q < 1$,
 as well as the corresponding
characteristic functions $\chi_n$.

Since
$$
\hat \mu_{AL}(\dot \Delta_n) \ \rightarrow \ 1 - q > 0,  \eqno(5.20)
$$
given any $\e > 0$ there is some $N \in \N$ such that
$ \forall n \geq m > N$ ,
$$
\parallel \chi_n - \chi_m \parallel^2 = \int_{{\cal A} / {\cal G}}
\left( \chi_n - \chi_m \right)^2 \ d \hat \mu_{AL}
= \hat \mu_{AL}(\dot \Delta_m) -
\hat \mu_{AL}(\dot \Delta_n)  < \e    \eqno(5.21)
$$
and the sequence $\{ \chi_n \}_{n=1}^\infty$ is Cauchy.
Suppose that it converges to some
$$
f \in {\cal C}L^2 \left(  {{\cal A} / {\cal G}} \right)
$$
which implies that $f$ is itself a
cylindrical function, $f = \tilde f \circ \hat \pi_{S^*_0}$
for some function $\tilde f$ on some $H_{S^*_0}$.

Consider now the
finitely generated subgroups $S^*_n = S^*[\hat \beta_1,...,
\hat \beta_n]$ used to define ${\dot \Delta}_n^{\{\epsilon
^{(q)}_i\}_{i=1}^n}$
and $\chi_n$.  For large enough $N$, no $\hat \beta_m$ for $m \geq N$
lies in $S^*_0$.  Thus, if $S^*{}'_m$, $m \geq N$, is the subgroup generated
by hoops in $S^*_m$ and hoops in $S^*_0$, $\chi_m(h)  = 0$ for any
homomorphism $h, \ [h]  \in H_{S^*{}'_n}$ such that $d_e(h(\hat \beta_N))
\leq \epsilon_N$. Let $R_m$ be the set of all such $[h]\in H_{S^*{}'_n}$.
Then $$
\eqalign{
|| \chi_m - f ||^2 &= \int _{H_{S^*{}'_n}} d \mu_{S^*{}'_n} |\chi_m - f|^2
\circ \pi^{-1}_{S^*{}'_n}
\cr
& \geq \int_{R_m} d \mu_{S^*{}'_n} |f|^2
\circ \pi^{-1}_{S^*{}'_n}
\cr
& =  s(\epsilon_N) \int_{S^*_0} d \mu_{S^*_0} |\tilde f|^2
} \eqno(5.22)
$$
so that $||\chi_m - f||^2$ is bounded away from zero unless $\tilde f$ is the
zero function.  However, if $f$ is the zero function then
$$
||\chi_m-f||^2 = ||\chi_m||^2 \geq q  \eqno(5.23)
$$
so that the Cauchy sequence
$\{ \chi_n \}_{n=1}^\infty$ does not converge in
${\cal C}L^2\left( {\cal A}/ {\cal G} \right)$ and
${\cal C}L^2\left( {\cal A}/ {\cal G} \right)$ is incomplete.

\vskip 1.5cm

\vbox{

{\bf Acknowledgements}

The authors would like to thank Abhay Ashtekar, John Baez,
Luis Barreira, Rodolfo Gambini,
Jerzy Lewandowski, Alan Rendall, Thomas Thiemann
 and Andrei T\"or\"ok
for their comments and for contributing to useful discussions
on the subject of measures on infinite dimensional spaces.
We would also like to thank the Gonzalez Courier Service of Syracuse,
New York for fast and faithful delivery.
This work was
supported in part by NSF grant PHY93-96246, by funds
provided by The Pennsylvania State University and JMM was also
 supported by  NATO grant 9/C/93/PO.

}

\vfill

\eject

{\bf References}

\vskip 0.5cm

\item{[1]} A. Ashtekar, ``Non-perturbative canonical quantum gravity"
(Notes prepared in collaboration with R.S. Tate), World Scientific,
Singapore, 1991

\item{[2]} A. Ashtekar and C. Isham, Class. Quant. Grav., {\bf 9}
(1992) 1433-85

\item{[3]} J. Glimm and A. Jaffe, ``Quantum physics", Springer-Verlag,
New York, 1987

\item{[4]} J. Baez, I. Segal, and Z. Zhou
``Introduction to algebraic and constructive quantum field theory,"
Princeton University Press, Princeton, 1992

\item{[5]} I.M. Gel'fand and N. Vilenkin, ``Generalized functions",
Vol. IV, New York, Academic Press, 1964

\item{[6]} A. Ashtekar and J. Lewandowski, ``Representation
theory of analytic holonomy $C^*$-algebras", preprint CGPG - 93/8-1.
To appear in Proceedings of the Conference ``Quantum Gravity and Knots",
edited by J. Baez (Oxford U.P.)

\item{[7]} J. Baez, ``Diffeomorphism-invariant generalized measures
on the space of connections modulo gauge transformations",
hep-th/9305045, To appear in Proceedings of the Conference
 ``Quantum Topology" edited by L. Crane and D. Yetter.

\item{[8]} Y. Yamasaki, ``Measures on infinite dimensional spaces", World
Scientific, Singapore, 1985

\item{[9]} W. Rudin, ``Functional analysis", McGraw-Hill, New York,
1973

\item{[10]} A. Rendall, Class. Quant. Grav. {\bf 10} (1993) 605-608

\item{[11]} W. Rudin, ``Real and complex analysis", McGraw-Hill, New York,
1987

\item{[12]} A.N. Kolmogorov and S.V. Fomin, ``Introductory Real Analysis",
Prentice-Hall Inc., 1970

\item{[13]} Yu. L. Dalecky and S.V. Fomin, ``Measures and differential
equations in infinite-dimensional space", Kluwer Academic Pub., Dordrecht,
1991

\vfill

\eject

\bye